\documentstyle[12pt]{article}
\advance\textheight by \topskip
\begin{document}
\thispagestyle{empty}
\begin{flushright}
{SU--ITP--94--3}\\
hep-th/9402115\\
February 13, 1994\\
\end{flushright}
 \vskip 1.3cm
\begin{center}
{\Large\bf TOPOLOGICAL DEFECTS \\
\vskip .9cm
AS SEEDS FOR ETERNAL INFLATION}\\
\vskip 2.5cm
  {\bf Andrei Linde}\footnote{On leave  from:  Lebedev
Physical Institute, Moscow, Russia.\
E-mail: linde@physics.stanford.edu},
\vskip 0.05cm
Department of Physics, Stanford University, Stanford, CA 94305-4060,
USA
\vskip .8cm
{\bf Dmitri Linde}\footnote{E-mail: dmitri@cco.caltech.edu}
\vskip 0.05cm
{California Institute of Technology,
Pasadena, CA 91125, USA}
\end{center}
\vfill
\eject
\hskip 1cm
\vskip 1cm

{\centerline{\large ABSTRACT}}
\begin{quotation}
\vskip -0.4cm
We investigate the  global structure of inflationary universe
both by analytical methods and by computer simulations of stochastic
processes
in the early Universe. We show that the global structure of
inflationary
universe depends crucially on the mechanism of inflation. In the
simplest
models of chaotic inflation  with the effective potentials  $\phi^n$
or
$e^{\alpha\phi}$ the Universe  looks like a sea of thermalized phase,
surrounding permanently self-reproducing inflationary domains.
On the other hand, in the theories where inflation may occur near a
local
extremum of the effective potential corresponding to a metastable
state, the
Universe looks like de Sitter space surrounding islands of
thermalized phase.
A similar picture appears even if the state $\phi = 0$ is unstable
but the
effective potential has a discrete symmetry, e.g. the symmetry $\phi
\to
=-\phi$. In this case the Universe becomes divided into domains
containing
different phases ($\eta$ or $-\eta$). These domains will be separated
from each
other by domain walls. However, unlike ordinary domain walls often
discussed in
the literature, these domain walls will inflate, and their thickness
will
exponentially grow.  In the theories with continuous symmetries
inflation will
generate exponentially expanding strings and monopoles  surrounded by
thermalized phase.  Inflating  topological defects will be stable,
and they
will unceasingly produce new  inflating topological defects.  This
means that
topological defects may play a role of indestructible seeds for
eternal
inflation.\\

 \end{quotation}
\vfill
\newpage

\section{Introduction}

Inflationary cosmology is gradually changing our point of view on the
global
structure of the Universe \cite{MyBook}. One of the most radical
changes  has
occurred when it was realized that in many versions of inflationary
theory the
process of inflation never ends. Originally this statement was shown
to be
correct for the old inflationary scenario \cite{b51}, and for the new
inflationary scenario \cite{b52,b62}. The main idea is that the field
near the
top   of the effective potential   does not move. Therefore if the
Universe
expands fast enough, there always will be enough space where the
field   stays
at the top (or occasionally jumps back to the top), and inflation
continues.
However,   this conclusion did not attract much attention. Old
inflation  did
not work anyway, and new inflation was also extremely problematic. It
was
plagued by the problem of initial conditions, and all its
semi-realistic
versions looked very complicated and not very natural \cite{MyBook}.

Chaotic inflation scenario \cite{Chaotic}  has brought two surprises.
First of
all, it was realized that inflation can occur even if there  was no
thermal
equilibrium in the early Universe, and even if  the effective
potential
$V(\phi)$ does not have any maximum at all, or if its maximum is not
sufficiently flat. In particular, chaotic inflation scenario can be
realized in
the theories with  potentials   ${m^2\over 2} \phi^2$,\  \
${\lambda\over 4}
\phi^4$,\  \  ${\lambda\over 4} ({\phi^2 - {m^2\over \lambda}})^2$,\
\ and \
$e^{\alpha\phi}$.
But the most surprising realization was that   inflation in these
theories also
goes on without end. Due to quantum fluctuations the scalar field
$\phi$ in
some parts of the Universe perpetually climbs to higher and higher
values of
its potential energy $V(\phi)$, until it approaches the Planck
density $M_P^4$.
The existence of this regime may  seem  counterintuitive. Indeed, the
probability that the field jumps up all the time is  very small.
However,
those rare domains where it happens continue growing exponentially,
much faster
than the domains with small $V(\phi)$. This scenario was called
``eternal
inflation''  \cite{b19}.

An important feature of this scenario was the existence of domains
where the
field $\phi$ may jump for a long time not far away from the Planck
density. In
these domains the Hubble constant is extremely large, $H \sim M_P$.
This
induces strong perturbations in all other scalar fields, which
eventually leads
to division of the Universe into exponentially large domains filled
with matter
with all possible types of symmetry breaking \cite{MyBook}, and maybe
even with
different types of compactification of space-time \cite{Zeln}. This
provides a
physical justification of the weak anthropic principle. Under certain
conditions, eternally inflating universe enters a  stationary regime,
where the
probability to find domains with given properties does not depend on
time
\cite{LM}. This is a considerable deviation of  inflationary
cosmology from the
standard big bang paradigm. A detailed discussion of this scenario
was given
recently in \cite{LLM}.

In the simplest versions of chaotic inflation scenario describing
only one
scalar field the Universe looks like a sea of a thermalized phase,
surrounding
islands of inflating space \cite{LLM}. A considerably different
picture appears
in the old inflationary theory, as well as in those versions of new
inflation
where the field $\phi$ can stay near the top of the effective
potential for a
long time, being in a kind of metastable state. For example, if the
probability
of formation of bubbles of the new phase in the old inflationary
universe
scenario is sufficiently small, the distance between previously
generated
bubbles grows exponentially before any new bubbles appear. Thus, the
new
bubbles appear far away from the old ones. In such a situation the
bubbles of a
new phase do not percolate; they always remain surrounded by de
Sitter space
\cite{GuthWeinb}. As similar conclusion was reached in \cite{b62a}
concerning
the structure of the Universe in the new inflationary universe
scenario. The
authors performed a computer simulation of inflation and of quantum
fluctuations in a simple theory with a potential which looked like a
step
function. It was equal to some positive constant $V(0)$ for
$\phi < \phi_0$, and   it was equal to zero for  $\phi > \phi_0$.
This
potential mimics many properties of realistic potentials used in the
new
inflation scenario. However, an important feature of this potential
was its
absolute flatness near $\phi = 0$, which effectively made the scalar
field near
$\phi = 0$ metastable. The conclusion of ref.  \cite{b62a} was that
in the new
inflationary universe scenario the Universe also consists of  islands
of
thermalized phase surrounded by de Sitter space.

In the present paper we will report the results of our investigation
of the
global structure of the Universe in the theories with   potentials of
the type
  ${\lambda\over 4} ({\phi^2 - {m^2\over \lambda}})^2$. Inflation  in
such
models may occur in two different regimes. If  it begins at $\phi >
M_P$, all
consequences will be the same as in the simple model  ${\lambda\over
4}
\phi^4$. This means that the inflationary domains   will look like
islands
surrounded by the thermalized phase.

On the other hand, for $m/\sqrt\lambda > M_P$, inflation may occur
near $\phi =
0$ as well, as in the new inflationary universe scenario. We will
argue that
the global structure of the Universe in this case will depend on the
properties
 of the theory. If we consider a theory of a real scalar field with a
discrete
symmetry $\phi \to - \phi$, the Universe will consist of islands of
thermalized phase with $\phi \sim m/\sqrt\lambda$.
However, if the scalar field has more than one component, for
example, if it is
a complex field $\phi = {1\over 2} (\phi_1+ i \phi_2)$,  then the
situation
will be different. The Universe will be filled by the thermalized
phase
containing inflating strings. In the $O(3)$-symmetric theory where
the scalar
field is  a vector $(\phi_1,\phi_2,\phi_3)$, thermalized phase will
surround
inflating monopoles. This means that topological defects may play an
extremely
important role in formation of the global structure of the Universe.

Investigation of this issue should help us to obtain a better
understanding of
a very interesting piece of physics which was missed in our previous
studies of
new inflation. Until very recently all experts in inflationary theory
believed
that primordial monopoles produced during inflation in the new
inflationary
scenario were effectively pointlike objects, which  did not inflate
themselves.
For example, in the first version of this scenario based on the
$SU(5)$
Coleman-Weinberg theory \cite{b16} the Hubble constant during
inflation was of
the order $10^{10}$ GeV, which is five orders of magnitude smaller
than the
mass of the $X$-boson $M_X \sim 10^{15}$ GeV. The size of a monopole
estimated
by $M_X^{-1}$   is five orders of magnitude smaller than the
curvature of the
Universe given by the size of the horizon $H^{-1}$. It seemed obvious
that such
monopoles simply could not know that the Universe is curved.

This conclusion finds an independent confirmation in the calculation
of the
probability of spontaneous creation of monopoles during inflation.
According to
\cite{GuthVil}, this probability is suppressed by   a factor of $\exp
(-2\pi
m/H)$, where $m$ is the monopole mass. In the model discussed above
this
factor is given by $\sim 10^{-10^6}$, which is negligibly small. This
result is
rather general. In all (or almost all) realistic models of inflation
the Hubble
constant $H$ at the end of inflation is smaller than $10^{14}$ GeV
\cite{MyBook}. Meanwhile most of the superheavy topological defects
that may
have interesting cosmological consequences  appear in the theories
with the
scale of spontaneous symmetry breaking $\eta \sim 10^{16}$ GeV, which
is at
least two orders of magnitude greater than $H$. The probability of
creation of
such topological defects  by the mechanism described in
\cite{GuthVil}  is
extremely small. Even if there were no barrier for production of such
objects,
their density would have been suppressed by a factor $\sim
\exp\Bigl(-{6\pi^2\eta^2\over H^2}\Bigr) \sim 10^{-10^5}$
\cite{Lyth}.\footnote{Superheavy topological defects can be created,
however,
during inflationary phase transitions  \cite{KL}.}

Despite all these considerations, in the present paper   (see also
\cite{L},
\cite{Vilenkin}) we will show  that in those theories where inflation
is
possible near a local maximum of the effective potential,
topological defects
 expand exponentially and can be copiously produced during inflation.
The main
reason can be explained as follows. In the cores of topological
defects the
scalar field $\phi$ always corresponds to the maximum of effective
potential.
When inflation begins, it makes the field $\phi$ almost homogeneous.
This
provides ideal conditions for inflation inside topological defects.
These
conditions  remain satisfied inside the topological defects even
after
inflation finishes outside of them. Moreover, as we are going to
argue, each
such  topological defect will create many other inflating topological
defects.
We have called such configurations  {\it fractal topological defects}
\cite{L}.

The paper is organized in the following way. In Section 2 we will
give a
description of inflation of domain walls at the level of classical
theory. In
Section 3 we will briefly describe this process with an account   of
quantum
fluctuations, and present the  results of our computer simulations of
this
process. In Section 4 we will describe similar processes in the case
of
inflating strings and monopoles. In Section 5 we will consider the
problem of
initial conditions for inflation near a local maximum of $V(\phi)$.
In the concluding Section 6 we will discuss our main results.

\section{Inflating domain walls}
To explain the basic idea of our work, we will begin with a
discussion of
inflating domain walls. The  Lagrangian of the  simplest  model where
such
walls may appear is given by
\begin{equation}\label{1}
L= {1\over 2} (\partial _\mu \phi )^2 - {\lambda\over 4} \Bigl(\phi^2
  - {m^2\over \lambda} \Bigr)^2  \ .
\end{equation}
Here  $\phi$ is a real scalar field.
Symmetry breaking in this model leads to formation of domains with
$\phi = \pm
\eta$, where $\eta = {m\over \sqrt \lambda}$. These domains are
divided by
domain walls     which interpolate between
the two minima. Neglecting gravitational effects, one can easily
obtain a
solution for a static domain wall  in the $yz$ plane:
\begin{equation}\label{2}
\phi = \eta\ {\tanh} \Bigl( \sqrt{\lambda\over 2}\ \eta x  \Bigr) \ .
\end{equation}

For small $\eta$  our neglect of gravitational effects is reasonable.
However,
the situation becomes more complicated if $\eta$ becomes comparable
to the
Planck mass $M_P$. The potential energy density in the center of the
wall
(\ref{2}) at $x = 0$ is equal to ${\lambda\over 4}  \eta ^4$, the
gradient
energy is also equal to ${\lambda\over 4}  \eta ^4$. This energy
density
remains almost constant  at $|x| \ll m^{-1} \equiv {1\over \sqrt
\lambda
\eta}$, and then it rapidly decreases.  Gravitational effects can be
neglected
if the Schwarzschild radius $r_g = {2M\over M_P^2}$ corresponding to
the
distribution of matter with energy density $\rho = {\lambda\over 2}
\eta ^4$
and radius $R \sim m^{-1}$ is much smaller than $R$. Here $M =
{4\pi\over 3}
\rho R^3$. This condition implies that gravitational effects can be
neglected
for $\eta \, \ll \, {3\over 2\pi}\, M_P$. In the opposite case,
\begin{equation}\label{3}
\eta \, {\
\lower-1.2pt\vbox{\hbox{\rlap{$>$}\lower5pt\vbox{\hbox{$\sim$}}}}\
}\, {3\over 2\pi}\, M_P  \ ,
\end{equation}
gravitational effects can be very significant.
A similar conclusion is valid for other topological defects as well.
For
example,   recently it was shown that magnetic monopoles in the
theory with the
scale of spontaneous symmetry breaking $\eta {\
\lower-1.2pt\vbox{\hbox{\rlap{$>$}\lower5pt\vbox{\hbox{$\sim$}}}}\
}M_P$ form
Reissner-Nordstr\"om black holes \cite{Lee}.

Now let us look at it from the point of view of inflationary theory.
Inflation
occurs at $\phi \ll \eta$ in the model (\ref{1}) if the curvature of
the
effective potential $V(\phi)$ at $\phi \ll \eta$ is much smaller than
$3 H^2$,
where $H = \sqrt{2\pi\lambda\over 3}\, {\eta^2\over M_P}$ is the
Hubble
constant supported by the effective potential \cite{MyBook}. This
gives
$m^2 \ll {2\pi\lambda \eta^4/ M_P^2}$, which leads to the condition
almost
exactly coinciding with (\ref{3}): $\eta \, \gg \,  M_P/\sqrt{2\pi}$.

This coincidence by itself does not mean that domain walls and
monopoles in the
theories with $\eta \, \gg \,  M_P/\sqrt{2\pi}$ will inflate. Indeed,
inflation
occurs only if the energy density is dominated by the vacuum energy.
As we have
seen, for the wall (\ref{2}) this was not the case:  gradient energy
density
for the solution (\ref{2}) near $x = 0$  is equal to the potential
energy
density. However, this is correct only after inflation and only if
gravitational effects are not taken into account.

At the initial stages of inflation the field $\phi$ is equal to zero.
Even if
originally there were any gradients of this field, they rapidly
become
exponentially small. Each time $\Delta t = H^{-1}$ new  perturbations
with the
 amplitude ${H/\sqrt{2} \pi}$ and the wavelength $\sim H^{-1}$ are
produced,
but their gradient energy density $\sim H^4$ is always much smaller
than
$V(\phi)$ for $V(\phi) \ll M^4_P$ \cite{LLM}.  Therefore originally
the vacuum
energy inside the walls dominated its gradient energy, and walls
could easily
expand. The reason why we did not understand this before is the same
as the
reason why we thought that the interior of the bubbles of the new
phase cannot
expand: We thought that the bubble walls during inflation were thin
from the
very beginning. Then we understood that this was wrong, and the new
inflationary scenario was proposed. Here we encounter the same
situation.
Domain walls, just as the bubble walls, originally were thick, and
they were
exponentially expanding.

\section{Self-reproduction of the Universe and fractal structure of
domain
walls}
Previous description was purely classical. Meanwhile quantum
fluctuations play
extremely important role in this scenario.

 The wavelengths of quantum
fluctuations of the scalar field $\phi$ grow exponentially
in the expanding Universe. When the wavelength of any
particular fluctuation becomes greater than $H^{-1}$, this
fluctuation stops oscillating, and its amplitude freezes at
some nonzero value $\delta\phi (x)$ because of the large
friction term $3H\dot{\phi}$ in the equation of motion of the field
$\phi$\@. The amplitude of this fluctuation then remains
almost unchanged for a very long time, whereas its
wavelength grows exponentially. Therefore, the appearance of
such a frozen fluctuation is equivalent to the appearance of
a classical field $\delta\phi (x)$ that does not vanish
after averaging over macroscopic intervals of space and
time.

One can visualize  fluctuations generated during the typical  time $
H^{-1}$ as
sinusoidal waves with average amplitude
\begin{equation}\label{c}
\delta \phi = {H \over   2 \pi} \  ~.
\end{equation}
and with a wavelength $\sim H^{-1}$.
Since phases of each wave  are
random,  the sum of all waves at a given point fluctuates and
experiences
Brownian jumps in all directions. As a result, the values of the
scalar
field in different points  become different from each other, and the
corresponding variance grows as
\begin{equation}\label{E23a0}
 {\left<\phi^2\right>} = {H^3\over 4\pi^2}\  t\ ,
\end{equation}
which means that dispersion grows as
$ \sqrt{\left<\phi^2\right>} = {H\over 2\pi}\ \sqrt{H\, t}$.

In general, the   Hubble constant $H$
strongly depends on the value of the scalar field $\phi$. However, we
consider
the  case when inflation occurs near a local maximum of the effective
potential
at $\phi = 0$. This gives  $H = \sqrt{2\pi \lambda\over
3}{\eta^2\over M_P}$,
and the average amplitude of fluctuations generated during the time
$H^{-1}$ is
given by
\begin{equation}\label{ca}
\delta \phi = \sqrt{\lambda\over 6\pi }{\eta^2\over M_P} = {m^2\over
\sqrt{6\pi\lambda}\, M_P}\  ~.
\end{equation}
These perturbations appear in the background of classically moving
field
$\phi$, which grows each time $H^{-1}$ by
\begin{equation}\label{cb}
\Delta \phi = {V'(\phi)
\over 3H^2 } = {\phi \lambda M_P^2\over 2\pi m^2} \  ~.
\end{equation}
Comparison of these two quantities shows that $\delta \phi > \Delta
\phi$ for
\begin{equation}\label{cc}
\phi <\phi^* \equiv  { m^4\, \sqrt 2  \over M_P^2\,
\lambda\sqrt\lambda}  \ .
\end{equation}

If from the very beginning the scalar field was sufficiently small,
$\phi \ll
\phi^*$, then the quantum jumps of the field $\phi$ could always
return the
field back to even smaller values of $\phi$. The field $\phi$ jumps
back only
in a half of domains with $\phi \ll \phi^*$. However,  this is quite
enough
since each typical time interval  $H^{-1}$ the total volume of such
domains
grows approximately $e^3/2 \sim 10$ times \cite{b62}.

Let us consider fluctuations near $\phi = 0$ in a more detailed way.
Suppose
that after the typical time $H^{-1}$ quantum fluctuations of the
scalar field
$\phi$ pushed it away from $\phi = 0$, and it acquired a positive
value $+
H/2\pi$ inside a domain of a size $H^{-1}$. During the next period of
time
$H^{-1}$ the original domain grows in size $e$ times, its volume
grows $e^3
\sim 20$ times. Therefore it becomes divided into $20$ domains of a
size of the
horizon $H^{-1}$. Evolution of the field inside each of them occurs
independently of the processes in the other domains (no-hair theorem
for de
Sitter space). In each of these domains the scalar field with a
probability
${1\over2}$ may jump back, or it may jump in the same direction.
However, these
jumps will occur  on the scale which is $e$ times smaller than the
length scale
of the previous fluctuation. In average those points which originally
jumped
to positive $\phi$ will remain positive, and the value of the field
$\phi$ at
these points will grow.

Suppose now that we paint white domains with positive $\phi$,   paint
grey
domains with negative $\phi$, and black -- the boundary between these
domains.
Then after the first step the domain will consist of two parts, one
is
homogeneously white, and another is homogeneously grey. After the
second
interval $H^{-1}$  the size of each domain will grow $e$ times. The
white
domain after expansion will contain some grey islands inside it, and
the grey
domain will contain some white islands.  These domains will be
separated by
black domain walls corresponding to $\phi = 0$. Only at the domain
walls does
the Universe return to its  state $\phi = 0$. Outside the walls the
field
$\phi$ always moves down to the minima of its effective potential.
After a
while, the Universe becomes divided into white and grey islands
separated from
each other by black domain walls. These domain walls still continue
expand
exponentially. Therefore qualitatively the picture we obtain is very
similar to
the one which emerges in the old inflationary scenario: The islands
of
thermalized phase are surrounded de Sitter space. However, the
physical reason
for this picture is somewhat different.

If the field $\phi$ is in a metastable state, or if it is in a state
of
equilibrium for a certain sufficiently large range of its values,
then the
bubbles of the new phase always appear surrounded by the old phase.
If the
decay rate of the old phase is small enough,   thermalized phase will
be always
surrounded by de Sitter space, even if the field can roll only in one
direction from its original position. On the other hand, the main
reason for
the existence of the domain structure of the  Universe in the model
under
consideration  is the possibility of the field $\phi$ falling down in
two
different directions from the maximum of the effective potential
$V(\phi)$. In
our model this was achieved due to the discrete symmetry $\phi \to
-\phi$ of
the effective potential. We should emphasize, however, that in fact
 we do not need exact or even
approximate symmetry. The same conclusions will remain valid for any
one-component   scalar field $\phi$ which has a potential $V(\phi)$
with a
sufficiently flat local maximum. This maximum can be at any point
$\phi_0$. The
flatness condition reads $V''(\phi_0) \ll H^2(\phi_0) = {8\pi
V(\phi_0)\over 3
M_P^2}$.

As we have mentioned already, the jumps of the field $\phi$ in our
model can
occasionally change its sign and create grey domains inside white
surroundings.
Simultaneously this forms new inflating domain walls.  These new
walls will be
formed only in those places where the scalar field is sufficiently
small for
the jumps with the change of the sign of the field $\phi$ to be
possible.
Therefore the new walls will be created predominantly near the old
ones (where
$\phi = 0$), thus forming a fractal domain wall structure.

As a part of our investigation, we made a series of computer
simulations of
this process in a two-dimensional slice of the Universe. All
calculations were
performed in comoving coordinates, which did not change
during the expansion of the Universe. In such coordinates,
expansion of the Universe results in an exponential shrinking of
wavelengths of perturbations. We  represent perturbations as
sinusoidal waves
in a two-dimensional universe,
\begin{equation}\label{c5}
\delta\phi(x,y)={H \sqrt u \over   \sqrt{2}\pi}\cdot
\sin\Bigl(H ~ e^{Ht}\, (x  \cos \theta_n + y  \sin \theta_n+
\alpha_n)\Bigr) \
{}.
\end{equation}
Here $u < 1$ is some small parameter which controls the time $\Delta
t = u
H^{-1}$ between two consequent steps of our simulations, $\theta_n$
and $
\alpha_n$ are random numbers. Equation (\ref{c5}) follows from the
corresponding equations of our paper \cite{LLM} in the case $H =
const$. This
is a very good approximation for describing inflation near the top of
the
effective potential. It fails in the thermalized regions, but
thermalized
regions do  not influence geometry of exponentially expanding part of
the
Universe at a distance greater than $H^{-1}$ from the boundary
between these
regions \cite{GuthWeinb}. This condition was satisfied during our
simulations.
Therefore we expect that our simulations correctly represent the
behavior of
the field not too far away from the top of the effective potential.
This is all
we need. One should not take too seriously the distribution of the
field $\phi$
in the regions with $V(\phi) \ll   V(0)$ in our figures. Fortunately,
this
distribution for $V(\phi) \ll   V(0)$ is of no interest for us.

We performed our  calculations using the grids
containing  $300 \times 300$ and $1000 \times 1000$ points.
At each step of our calculations we added to the previous
distribution of the
field the wave (\ref{c5}) and also took into account the classical
drift  of
the field by
\begin{equation}\label{c6}
\Delta\phi(x,y)=-\, {u V'(\phi)\over 3 H^2} \ \, .
\end{equation}

A more detailed description of our method can be found  in
\cite{LLM}. Here we
will just briefly describe our results, which are shown in Fig. 1.

As we mentioned above, we paint black the regions corresponding to
the domain
walls. However, in our figures we included into the definition of a
domain wall
all points where $|\phi| < \phi^*$. Thus, the points in white and
grey area
($\phi > \phi^*$, and $\phi <  \phi^*$) practically never change
their color,
since at $|\phi| > \phi^*$ the amplitude of quantum jumps $\delta
\phi$
typically is much smaller than the
classical drift $\Delta \phi$. Therefore one can consider these
regions as the
regions containing thermalized matter. We begin our simulations in a
domain of
a typical size $H^{-1}$  with a field $\phi \ll \phi^*$. As we see,
after a few
steps white and grey islands appear inside the black area, Fig. 1a.
Then new
and new islands become formed, Figs. 1b--1d. The fractal structure of
the
domain walls is obvious from these simulations. These simulations are
similar
to those performed in   an important paper by Aryal and Vilenkin
\cite{b62a}.
The   method used in the present work allows us to reveal the
physical nature of  the exponentially expanding phase. This phase
corresponds
to expanding domain walls dividing regions filled by {\it different}
phases.
Note that there are no  black  walls  which separate white regions
from white
regions. (Such walls would exist in the old inflationary scenario.)
This
demonstrates an important role   topological defects can play in
inflationary
cosmology: they can determine the global structure of the Universe.
This
suggests also that in the models with different types of topological
defects,
the global structure of the Universe may   look different.

\section{Inflating strings and monopoles}
 We will consider now more complicated  models where instead of  a
discrete
symmetry $\phi \to - \phi$ we have a continuous symmetry. For
example, instead
of the model (\ref{1}) describing a real scalar field one can
consider a model
\begin{equation}\label{5}
L= \partial _\mu \phi^*\,  \partial _\mu \phi  - {\lambda}
\left(\phi^*\phi
  - {\eta^2\over 2 } \right)^2  \ ,
\end{equation}
where  $\phi$ is a complex scalar field, $\phi = {1\over \sqrt
2}(\phi_1 + i
\phi_2)$.
Spontaneous  breaking of the $U(1)$ symmetry in this theory may
produce global
cosmic strings. Each string contains  a line with $\phi = 0$. Outside
this line
the absolute value of the field $\phi$ increases and asymptotically
approaches
the limiting value  $\sqrt{\phi_1^2 + \phi_2^2} = \eta$. This string
will be
topologically stable if the isotopic vector $(\phi_1(x), \phi_2(x))$
rotates by
$2n\pi$ when the point $x$ takes a closed path  around the string.

The (global) monopole solutions for the first time appear in the
theory with
$O(3)$ symmetry,
\begin{equation}\label{6}
L= {1\over 2}  (\partial _\mu\vec \phi )^2 -
{\lambda \over 4} ( {\vec \phi}^2 - {{\eta ^ 2}} ) ^2
  \ ,
\end{equation}
where $\vec \phi$ is a vector $(\phi_1, \phi_2, \phi_3)$.  The
simplest
monopole configuration contains a point $x = 0$ with $\phi(0) = 0$
surrounded
by the scalar field $\vec \phi(x) \propto \vec x$. Asymptotically
this field
approaches regime with $\vec \phi^2(x) = \eta^2$.

The basic feature of all topological defects including strings and
monopoles is
the existence of the points where $\phi = 0$. Effective potential has
an
extremum at $\phi = 0$, and if the curvature of the effective
potential is
smaller than $H^2 = {8\pi V(0)\over 3M^2_P}$, space around the points
with
$\phi = 0$ will expand exponentially, just as in the domain wall case
considered above.

Now we can add gauge fields. We begin with the Higgs model, which is
a direct
generalization of the model (\ref{5}):
\begin{equation}\label{5a}
L= D _\mu \phi^*\,  D _\mu \phi  -  {1\over 4} F _{\mu \nu} F^{\mu
\nu}   -
{\lambda} \left(\phi^*\phi
  - {\eta^2 \over 2 }\right)^2  \ .
\end{equation}
Here $D _\mu$ is a covariant derivative of the scalar field, which in
this
case  is given by $\partial _\mu - i e A_\mu$. In this model strings
of the
scalar field contain  magnetic flux $\Phi = 2\pi/e$. This flux is
localized
near the center of the string with $\phi(x) = 0$, for the reason that
the
vector field becomes heavy at  large $\phi$, see e.g.
\cite{Kirzhnits}.
However, if inflation  takes place inside the string, then the field
$\phi$
becomes vanishingly small not only at the central line with $\phi(x)
= 0$, but
even exponentially far away from it. In such a situation the flux of
magnetic
field will not be confined near the center of the string. The
thickness of the
flux will grow together with the growth of the Universe. Since the
total flux
of magnetic field inside the string is conserved, its strength will
decrease
exponentially,
and very soon its effect on the  string expansion will become
negligibly small.
Therefore vector fields will not prevent inflation of strings.

The final step is to consider  magnetic monopoles. With this purpose
one can
add non-Abelian gauge fields $A^a_\mu$ to the $O(3)$ symmetric theory
(\ref{6}):
\begin{equation}\label{7}
L= {1\over 2}  |D _\mu {\vec \phi} | ^2  - {1\over 4} F^a _{\mu \nu}
F^{a \mu
\nu}-
{\lambda \over 4} ( {\vec \phi}^2 - {{\eta ^ 2}}  ) ^2  \ .
\end{equation}
Global monopoles of the theory (\ref{5}) become magnetic monopoles in
the
theory (\ref{5a}). They also have $\phi = 0$ in the center.  Vector
fields in
the center of the monopole are massless ($g\phi = 0$). During
inflation these
fields exponentially decrease, and therefore they do not affect
inflation of
the monopoles.

We should emphasize that even though the field $\phi$ around the
monopole
during inflation is very small, its topological charge is well
defined,  it
cannot change and it cannot annihilate with the charge of other
monopoles as
soon as the radius of the monopole becomes greater than $H^{-1}$.
However, an
opposite process is possible. Just as domain walls can be easily
produced by
quantum fluctuations near other inflating domain walls, pairs of
monopoles can
be produced in the vicinity of an inflationary monopole. The distance
between
these monopoles grow exponentially, but the new monopoles will appear
in the
vicinity of each of them. We will show how it happens using computer
simulations of this process.

Note that in the simple models discussed above inflation of monopoles
occurs
only if spontaneous symmetry breaking is extremely strong, $\eta {\
\lower-1.2pt\vbox{\hbox{\rlap{$>$}\lower5pt\vbox{\hbox{$\sim$}}}}\
}M_P$.
However, this is not a necessary condition. Our arguments remain
valid for
all  models where the curvature of the effective potential near $\phi
= 0$ is
smaller than the Hubble constant supported by $V(0)$. This condition
is
satisfied by all models which were originally proposed for the
realization of
the new inflationary universe scenario.
In particular, the monopoles in the $SU(5)$ Coleman-Weinberg theory
also should
expand exponentially. The reason why we thought that this is
impossible was
explained in the Introduction: The Hubble constant $H$ during
inflation in the
$SU(5)$ Coleman-Weinberg theory is much smaller than the mass of the
vector
field $M_X$, which is usually related to the size of the monopole.
However, this argument is misleading. The effective mass of the
vector field
$M_X  \sim g\eta \sim 10^{15}$ GeV can determine  effective size of
the
monopole only {\it after} inflation.   Effective mass of the vector
field
$M_X(\phi) \sim g\phi$ is always equal to zero in the center of the
monopole.
Once inflation begins in a domain of a size $O(H^{-1})$ around the
center of
the monopole, it expels vectors fields away from the center and does
not allow
them to penetrate back as far as inflation continues.

Of course, one may argue that there is no much reason to consider
inflation
generated by magnetic monopoles. If the inflaton field is not a
gauge singlet,
the density perturbations produced after inflation typically are too
large
\cite{MyBook}.
However, there may be many different stages of inflation, and the
last one can
be driven by a different mechanism.
The main problem is how to obtain good initial conditions for the
first stage
of inflation and (if possible) how to make it eternal. Here
topological
defects may be of some help.

An interesting feature of this scenario is that inflation of
monopoles is
eternal for
purely classical (topological) reasons \cite{L,Vilenkin}. There is
the only way
for a monopole to stop inflating.  Even though we have estimated the
amplitude
of quantum fluctuations around the monopole to be very small,
eventually at
some moment this amplitude may appear to be much larger than its
typical value
$H/2\pi$. The probability of large jumps of the scalar field $\phi$
is
exponentially small \cite{LLM}, but small probabilities can
accumulate when we
are speaking about eternity. If  the gradients of the classical field
$\phi$
become sufficiently large because of the large fluctuation
$\delta\phi$, the
monopole may stop inflating. However, the probability of this event
is much
smaller than the probability of the monopole pair creation in the
vicinity of
an expanding monopole. Therefore quantum fluctuations which may kill
inflation
of the    monopole simultaneously create many new inflationary
monopoles.
Moreover, even if quantum fluctuation can terminate inflation of a
monopole,
they certainly cannot do the same for inflating strings and domain
walls.

We have performed computer simulations illustrating some of these
issues.
Our simulations were two-dimensional, and analogs of the monopoles
were the
centers of the strings in the model (\ref{5}). The centers of the
monopoles
should correspond to the points where $\phi_1 = \phi_2 = 0$. There
are three
series of figures in our  simulations. Fig. 2 shows the distribution
of
potential energy density in the two-dimensional Universe. In the
beginning
potential energy density is equal to $V(0) = {m^4\over 4\lambda}$ in
the whole
domain of initial size $H^{-1}$. After a few steps of expansion the
surface
$V(\phi(x,y))$ shown in Fig. 2a, bends a little, but still the  value
of the
effective potential does not differ much from $V(0)$. (The box
(x,y,V) is not
shown in this figure.) Later it decreases everywhere except for some
points
where it remains equal to $V(0)$.  These points are the peaks of the
mountains
surrounded by the thermalized phase in Fig. 2. In the beginning we
see just a
few such mountains, Fig. 2b, but then they split and form new
mountains
separated from others by the thermalized phase, Figs. 2c, 2d. Note
that all
these mountains have equal height. It is instructive to compare this
picture
with a typical distribution obtained in the chaotic inflation
scenario with the
potential $V(\phi) = {m^2\over 2} \phi^2$, Fig. 3. In this case
mountains are
also separated by the thermalized phase, but their height can be as
large as
$M_P^4$.

Our calculations have been performed with several different sets of
parameters.
 The
results shown in Fig. 2 correspond  to $m = 0.3\, M_P$, $\lambda =
0.09$, $\eta
= M_P$.  Inflationary condition $m^2 \ll 3H^2$ is satisfied for these
values of
 parameters. Of course, these parameters are far from their values in
realistic models. Still they should give us a qualitatively correct
picture of
the process.

It is very tempting to identify the peaks of the  mountains  shown at
Fig. 2
with monopoles. However,  most of the     mountains correspond to
topologically trivial field configurations. Moreover, most of them do
not even
have   $\phi = 0$ in the center. Indeed, the only condition which is
necessary
for the self-reproduction of inflationary domains with large
$V(\phi)$ is that
the absolute value of the field $\phi$ should be smaller than
$\phi^*$
(\ref{cc}). This means that $V(\phi)$ at the peaks of the mountains
is
somewhere in the interval  $V(\phi^*) \leq V(\phi) \leq V(0)$.
Typically this
means that $V(\phi)$ on the peaks of the mountains is very close to
$V(0)$, but
it may be slightly different from $V(0)$.  Thus one should not
overemphasize
the role
of topological defects in the eternal process of self-reproduction of
the
Universe. This process can occur without topological defects as well.
Still the
possibility of exponential expansion and self-reproduction of
topological
defects adds some new dimension to this theory.

The field $\phi$ should make many jumps back from $\phi^*$ to $\phi =
0$.
Consequently, the number of the monopoles produced due to these jumps
will be
suppressed by a factor $\sim \exp(-4\pi^2\phi^2/H^2)$. Monopoles will
be
copiously produced in this scenario, but only   near the points where
the field
$\phi$ is sufficiently small,  $|\phi_1|, |\phi_2|   {\
\lower-1.2pt\vbox{\hbox{\rlap{$<$}\lower5pt\vbox{\hbox{$\sim$}}}}\ }
H/2\pi$,
in
particular, near other monopoles.

 In order to identify those mountains which correspond to monopoles
we
performed another series of computer simulations.  We used color to
show the
direction of the vector $(\phi_1(x), \phi_2(x))$ in the isotopic
space. Namely,
we used white color  if this vector was looking in the direction
$(1,0)$ (i.e.
positive $\phi_0$ and vanishing $\phi_2$), and then we gradually
increased the
level of darkness as the vector  $(\phi_1(x), \phi_2(x))$ rotated by
the angle
approaching $2\pi$. The point $2\pi$ for obvious reasons corresponds
to a
discontinuity; the color is either white or black depending on the
way we
approach it. This discontinuity does not imply existence of any
physical
singularity.  However, this color map allows us to  identify the
monopoles as
the points where the boundary lines between black and white end  in a
grey
area.

 Fig. 4 shows the distribution of the direction of the  vector
$(\phi_1(x),
\phi_2(x))$ using this color map. As we can see,   monopoles  are
created in
this process, and   their distribution indeed looks like a fractal,
which
becomes more and more complicated in the course of time.  (For the
attentive
reader: there are eight monopoles in Fig. 4a and thirteen monopoles
in Fig.
4b.)  However, if we impose these   pictures on the distribution of
the energy
density $V(\phi)$, we will see that some mountains correspond to
monopoles, and
some do not, see Fig. 5.  The stage of the process shown in  Fig. 5c
corresponds to  the  field distribution in   Fig. 4a. The first
monopole can be
seen in the upper right part of Fig. 5a.

As we already mentioned,   the  centers of inflationary domains in
Fig. 2   do
not form walls surrounding the thermalized phase. On the contrary,
inflating
domains are surrounded by the thermalized phase. The reason for this
behavior
in the simplest versions of chaotic inflation scenario  can be easily
understood.
Nothing prevents the field $\phi$ at each particular point to roll
down to the
minimum of the effective potential. Only very rarely the field $\phi$
jumps
against the classical flow down. Those  rare points where this
happens   form
the peaks of mountains in Fig. 3. After a sufficiently large time
these peaks
become surrounded by the thermalized phase.

 As we already mentioned, in the situation where the state $\phi = 0$
is
metastable with a sufficiently large lifetime we would  encounter an
opposite
regime. Independently of all topological considerations we would
obtain islands
of thermalized phase surrounded by de Sitter space. Is there any
strict
boundary between these two regimes?  Is it possible that topological
defects
will prevent rolling of the field $\phi $ down to the minimum of
$V(\phi)$ in a
 considerable part of space  even in the situations where the state
$\phi = 0$
is unstable?

One can get some insight by a more detailed investigation  of the
shape of
domain walls (generalization to monopoles is straightforward) by
using a slight
extension of  the method of  ref. \cite{Vilenkin}. Let us assume that
the field
$\phi$ initially is very small, $\phi \ll \eta$, and its
configuration
$\phi(x,0)$ is sufficiently smooth. Here   $x$ is a comoving
coordinate  of the
point we consider. We will assume also that near the center of a
domain wall
one can write $\phi( x,0)$ in the first approximation as $c x$, where
$c$ is
some small constant.  In this case the amplitude of the scalar field
at each
particular point will grow exponentially  \cite{MyBook},
\begin{equation}\label{v1}
\phi(x,t) \approx c\, x\, \exp \left({m^2\, t\over 3H}\right) \ ,
\end{equation}
Let us  write this  equation   in terms of the physical distance $X =
x \,
e^{Ht}$:
\begin{equation}\label{v2}
\phi  \approx c X \exp \left[- \left(H- {m^2\over 3H}\right) t\right]
\ .
\end{equation}
This equation means that at a physical distance
$X  \sim  \exp \left[\left(H- {m^2\over 3H}\right) t\right]$
from the center of the topological defect the value of the field
$\phi$ does
not change in time. In other words, inflation stretches the domain
wall without
changing its shape at small $\phi$. However, the Universe stretches
domain
walls in the $x$-direction more slowly that it stretches itself. In
the
comoving coordinates the thickness of the wall exponentially
decrease. Indeed,
one can easily see that the value of the field (\ref{v1}) does not
change for
\begin{equation}\label{v3}
x  \sim  \exp \left(-{m^2 t\over 3H}\right)  \ .
\end{equation}
 This makes it easier to understand the difference between the
topological
structure of the Universe in the old inflation scenario and in the
new one. In
the old inflation scenario de Sitter phase decays due to spontaneous
appearance
 of holes inside it, which leads to formation of islands of
thermalized phase
surrounded by de Sitter space. In our scenario the state $\phi = 0$
is
unstable, and all space has a tendency to go to the thermalized
phase.
Inflation still continues near the regions with $\phi = 0$, but the
comoving
size of these regions exponentially decreases in some directions. In
the case
of domain walls this is not very important; they surround domains of
thermalized phase for topological reasons.\footnote{The possibility
of
percolation of thermalized domains in a three-dimensional space
remains an open
question \cite{GuthWeinb,MM}.} On the other hand,  shrinking   (in
the comoving coordinates) strings and monopoles gradually become
surrounded by
the thermalized phase. This picture is consistent with   the results
of our
calculations.

Note, however, that our last argument was based on the assumption
that the
effective potential is quadratic near $\phi = 0$. Meanwhile in the
Coleman-Weinberg model the effective potential near the maximum looks
like
$V(0) - {\lambda\over 4} \phi^4$. Behavior of  domain walls in this
theory is
more complicated. At small $\phi$ the field decreases more slowly.
This changes
the shape of the domain wall, making it more flat near $\phi = 0$,
which more
closely resembles the situation in the old inflation scenario. On the
other
hand, one can argue that due to quantum fluctuations the field $\phi$
spends
most of the time at $\phi$  greater than $H/2\pi$, and therefore in
average is
rolls down at least as fast as the field in the theory (\ref{1}) with
$m^2
\sim  \lambda H^2$. Therefore we expect that our conclusions
concerning the
global structure of the Universe will remain qualitatively correct
for the
theories with the effective potentials  $\sim V(0) - {\lambda\over 4}
\phi^4$.
However, this subject clearly requires further investigation.

\section{The problem of initial conditions}
The possibility of inflation of topological defects can lead to some
improvement with the problem of initial conditions in the models
where
inflation occurs near a local maximum of $V(\phi)$. Initially the
models of
that type  were introduced in the context of the new inflationary
universe
scenario \cite{b16}. The basic assumption of old and new inflation
was that
inflation begins   in a state of thermal equilibrium at $\phi = 0$.
This idea
was not particularly successful, and no realistic versions of new
inflation
were suggested so far.  Still it is possible for inflation to begin
at $\phi =
0$ in the context of chaotic inflation scenario if  for some reason
the scalar
field appears near the top of the effective potential inside a domain
of a size
greater than $H^{-1}$. But is it possible to achieve it in a natural
way?

In order to analyse this question let us imagine that we are
witnessing the
moment of the Universe creation (``Planck time''), when the first
domain of
classical space-time with the Planck energy density $M_P^4$ emerged
from the
space-time foam.  It seems extremely unlikely that this first domain
is
infinite from the very beginning. In this case we would face the
horizon
problem: How was it possible for the same event (the appearance of
matter with
the Planck density) to be correlated in infinitely many causally
disconnected
domains?

The only natural length scale in   general relativity theory is the
Planck
length $M_P^{-1}$. Therefore the most natural assumption is that the
initial
domain has the Planck length.\footnote{ Of course, this size might be
much
larger if there was a preceding stage of evolution of the Universe,
for example
something like stringy pre-inflation \cite{Veneziano}. This
possibility is
extremely interesting, but its discussion is outside the scope of the
present
paper.}
If inflation of this domain does not begin immediately after that,
there is a
good chance that such a domain will momentarily collapse within the
time
$M_P^{-1}$. This is definitely the case if this domain locally looks
like a
part of a closed universe, but even if the domain looks like a part
of an open
universe of a size $M_P^{-1}$ immersed into   space-time foam, the
only obvious
way for it to avoid collapse and to evolve into a large homogeneous
universe
would be to begin inflation instantaneously.

This is not a problem at all for the simplest versions of chaotic
inflation,
where inflation can easily begin at $V(\phi) \sim M_P^4$
\cite{MyBook}.
However,  in all   models where inflation  occurs near the top of the
effective
potential, the value of $V(0)$ appears to be at least ten orders of
magnitude
smaller than the Planck density, and typically it is even much
smaller than
that.  Inflation in such models can begin only at a much later stage
of the
evolution of the Universe, at a time $t \sim M_P/\sqrt{V(0)}  {\
\lower-1.2pt\vbox{\hbox{\rlap{$>$}\lower5pt\vbox{\hbox{$\sim$}}}}\
}10^5
M_P^{-1}$. The size of   initial domain of inflationary universe at
that time
should be greater than $\Delta x \sim H^{-1} \sim M_P/\sqrt{V(0)}$.
Suppose for
simplicity that the Universe from the very beginning was dominated by
ultrarelativistic matter. Then its scale factor expanded as
$\rho^{-4}$, where
$\rho$ is the energy density at the pre-inflationary stage. Therefore
at the
Planck time the size of the part of the Universe which later evolved
into
inflationary domain was not  $  M_P/\sqrt{V(0)}$, but somewhat
smaller: $\Delta
x \sim V^{-1/4}(0)$. This whole scenario can work only if at the
Planck time
the domain of this size was sufficiently homogeneous, ${\delta \rho
\over \rho}
\ll 1$. However, at the Planck time this domain   consisted of
$M_P^3\,
V^{-3/4}(0)$ domains of a Planck size, and energy density in each of
them was
absolutely uncorrelated with the energy density in other domains.
Therefore
{\it a priori} one could expect changes of density $\delta \rho \sim
\rho$ when
going from one    causally disconnected parts of the Universe of a
size
$M_P^{-1}$ to another. Simple combinatorial analysis suggests that
the
probability of formation of  a reasonably homogeneous part of the
Universe of a
size  $\Delta x \sim V^{-1/4}(0)$ at the Planck time is suppressed by
the
exponential factor
\begin{equation}\label{d1}
P \sim \exp\Bigl(-{C\, M_P^3\over V^{3/4}(0)}\Bigr) \ ,
\end{equation}
where $C = O(1)$.
To get a numerical estimate, one can take $V(0) \sim 10^{-10} M_P^4$.
This
gives $P  {\
\lower-1.2pt\vbox{\hbox{\rlap{$<$}\lower5pt\vbox{\hbox{$\sim$}}}}\
} 10^{-10^{7}}$. For the original $SU(5)$ Coleman-Weinberg model this
number is
even much smaller.    Note that this estimate is very similar to the
estimate
of the probability of a direct quantum creation of inflationary
universe with
the vacuum energy density $V(\phi)$ \cite{Creation},
\begin{equation}\label{d2}
P \sim \exp\Bigl(-{3M^4_P\over 8V(\phi)}\Bigr) \ ,
\end{equation}
which gives even smaller value of the probability of inflation at
$\phi =
0$ than eq. (\ref{d1}). Meanwhile this equation   tells us that there
is no
suppression of probability of chaotic inflation with $V(\phi) \sim
M_P^4$.

One of the differences between these two estimates is that  eq.
(\ref{d1})
still does not guarantee that the homogeneous part of the Universe
will
inflate.   Inflation begins near the local maximum of the effective
potential
only if the  field $\phi$ in this domain   appears in a state with
$\phi \ll
\eta$, and is sufficiently homogeneous. Meanwhile in the theory
(\ref{1}) the
field $\phi$ initially can take any value in the interval from  $ -
\lambda^{-1/4} M_P$ to $+ \lambda^{-1/4} M_P$.  In realistic models
with
$\lambda \sim 10^{-12}$ this means that the typical initial value of
$\phi$
would be of the order of $10^3 M_P$. Then it will participate in
inflation and
roll  down to $\phi = \eta$, just as in the simplest versions of
chaotic
inflation scenario. The probability  to obtain a domain containing
homogeneous field $\phi \ll \eta$ (assuming that $\eta \ll 10^3 M_P$)
will be
even smaller than $\exp\Bigl(-{C\, M_P^3\over V^{3/4}(0)}\Bigr)$. At
this stage
topology may help  \cite{Vilenkin}. Once we have a sufficiently large
and
homogeneous domain, it is most probable that the field $\phi$ in the
model
(\ref{1}) will take both positive and negative values in its
different parts.
Consequently, there will be domain walls. Since in this model domain
walls can
be stretched by inflation, they will be even more easily stretched at
the
pre-inflationary stage, because at that time the Hubble constant was
even
greater. This naturally creates good conditions for inflation inside
the domain
wall. However, equations (\ref{d1}) and  (\ref{d2}) clearly indicate
that the
probability to obtain inflation beginning at large $\phi$ is much
better.

Does this mean that we should abandon the idea of chaotic inflation
near the
local maximum of  effective potential? In our opinion, this would be
incorrect.
First of all, it might happen that in a future theory of   elementary
particles
inflation cannot occur anywhere else except for a local maximum of
$V(\phi)$.
Still it  will be much better than no inflation at all.
On the other hand, there exist several different ways to create good
conditions
for inflation at $\phi = 0$. The simplest way is to add to the theory
some
heavy field $\Phi$ with a simple effective potential for which
inflation may
begin at  $V(\Phi) \sim M_P^4$. This stage of inflation  initiated by
the field
$\Phi$  will force the field
$\phi$ to jump to the top of the effective potential $V(\phi)$ at
least in some
part of the Universe.  Initially the part of the volume of the
Universe where
the field $\phi$ stays at the top of the effective potential will be
relatively
small, but later these regions will become increasingly important,
since they
will eternally inflate \cite{LLM}.

Another way is to   consider potentials of the new inflationary type
in the
context of the Brans-Dicke theory. In this case the Planck mass
depends on the
value of the Brans-Dicke field $\Phi$, and the condition $V(\phi)
\sim
M_P^4(\Phi)$ can be satisfied at the local maximum of $V(\phi)$
\cite{Hybrid,Bellido}.

There is also another interesting possibility \cite{Mijic,LLM}. The
wave
function of the Universe should describe all possible initial
conditions and
all possible outcomes. However, we are interested only in the
conditional
probability to obtain particular observational data under  an obvious
but very
nontrivial condition of our own existence. There may be many branches
of the
wave function of the Universe which may
seem natural from the point of view of initial conditions, but most
of them
describe the Universe where intelligent observers cannot live. In our
calculation of the probability (\ref{d1}) we simply counted all
trajectories,
even those which correspond to   ``virtual'' universes collapsing
within the
Planck time. But why should we count them? Perhaps we should see
where most of
the observers can live, and we should call the corresponding
trajectories
``typical''.  There will be many problems with such approach, in
particular the
problem of introducing a proper measure on the set of all such
trajectories
\cite{Bellido}. However,  it seems plausible that  with any
reasonable choice
of measure the trajectories corresponding to eternal inflation will
always win
being compared to the trajectories which do not possess this
property.

The only real problem appears if we should compare many different
possibilities
corresponding to different realizations of the eternal inflation
scenario. In
this case one   should take into account that it is much more
difficult for
inflation to begin  at $V(\phi) \ll M_P^4$ than at $V(\phi) \sim
M_P^4$.

\section{Discussion}

When we began this investigation, our main purpose was to study the
difference
between the global structure of the Universe in the models of two
different
classes: those models where inflation
occurs near a local maximum of the effective potential and those
models where
inflation begins at large $\phi$, outside the equilibrium.  However,
during our
work we recognized that some other important features of inflation in
the
models of the first class
theories  have not been properly analysed. For more than ten years we
knew that
inflation solves the primordial monopole problem, but we did not know
that
monopoles and other topological defects can inflate.

Now it appears that under certain conditions they do inflate, and
their
inflation never ends \cite{L,Vilenkin}.  According to this scenario,
our part
of the Universe could be formed from what initially was an interior
of an
inflating topological defect.  The first attempt to investigate this
question
was made in our paper \cite{LLM}, where we have shown that in
accordance to the
most natural realization of the ``natural inflation'' scenario
\cite{Natural}
we should live in the remnants of  an inflating domain wall. Now we
understand
that this situation is much more general.

The structure of space-time near inflating topological defects is
very
complicated; it should be studied by the methods developed in
\cite{Berezin}
for description of a bubble of de Sitter space immersed into vacuum
with
vanishing energy density.  Depending on initial conditions, many
possible
configurations may appear. For example, an inflating  monopole may
look from
outside  like a small magnetically charged Reissner-Nordstr\"om black
hole
\cite{Lee}. However, it will contain  a part of exponentially
expanding space
inside it. This will be  a wormhole configuration similar to those
studied in
\cite{Berezin}--\cite{Tkachev}.

At the quantum level the situation becomes even more interesting.
Fluctuations
of the field $\phi$ near the center of a monopole are strong enough
to create
new regions of space with $\phi = 0$, some of which will become
monopoles.
After a while, the distance between these monopoles becomes
exponentially
large, so that they cannot annihilate. This process of
monopole-antimonopole
pair creation produces a fractal structure consisting of monopoles
created in
the vicinity of other monopoles.

One of the original motivations for the development of inflationary
cosmology
was a desire to get rid of primordial magnetic monopoles and
dangerous domain
walls produced in the theories with spontaneous breaking of discrete
symmetries.  For a long time topological defects and inflation were
opposed to
each other as two almost incompatible sources of density
perturbations in the
early Universe. Now we see that the interplay between inflationary
theory and
the theory of topological defects can be very constructive.
According to our
scenario, inflation can produce   inflating topological defects which
in their
turn can serve as  seeds for eternal inflation.

We are very grateful to  Victor Berezin, Valery Frolov, Renata
Kallosh,  Arthur
Mezhlumian, Igor Tkachev  and especially to Alex Vilenkin for
valuable
discussions.   This work was supported in part  by NSF grant
PHY-8612280.

\vfill
\newpage

\section*{Figure Captions}
\begin{description}

\item[Fig. 1] The domain structure of the Universe in the theory
(\ref{1}) with
  spontaneously broken discrete symmetry $\phi \to - \phi$.

\item[Fig. 2] Energy density distribution during inflation in the
theory
(\ref{5}).

\item[Fig. 3] Energy density distribution during inflation in the
simplest
chaotic inflation model with the effective potential ${m^2\over 2}
\phi^2$.

\item[Fig. 4] Distribution of the field $\phi = {1\over \sqrt 2}
(\phi_1 +
i\phi_2)$ during inflation in the theory (\ref{5}).

\item[Fig. 5] These figures show simultaneously the energy density of
the field
$\phi$ in the theory  (\ref{5}), and  its direction in the isotopic
space.

\end{description}
\end{document}